\documentclass[conference]{IEEEtran}
\IEEEoverridecommandlockouts
\usepackage{cite}
\usepackage{amsmath,amssymb,amsfonts}
\usepackage{algorithmic}
\usepackage{graphicx}
\usepackage{textcomp}
\usepackage{pgfplots}
\pgfplotsset{compat=1.18}

\usepackage{tabularx}
\usepackage{booktabs}
\usepackage{comment}
\usepackage{hyperref} % for emphasizing the cross-references
\usepackage{url}

\usepackage[autostyle]{csquotes}
\MakeOuterQuote{"} % Correct US quotation mark

\usetikzlibrary{shapes.geometric, arrows.meta, positioning} 
\tikzset{
    terminator/.style={rectangle, draw, text centered, rounded corners, minimum height=2em},
    process/.style={rectangle, draw, text centered, minimum height=2em},
    decision/.style={diamond, draw, text centered, minimum height=2em, minimum width=6em, aspect=2},
    data/.style={trapezium, draw, text centered, trapezium left angle=60, trapezium right angle=120, minimum height=2em},
    connector/.style={draw, -Latex}
}

\usepackage{physics}
\usepackage[dvipsnames]{xcolor}
\definecolor{supergreen}{HTML}{009900}

\def\BibTeX{{\rm B\kern-.05em{\sc i\kern-.025em b}\kern-.08em
    T\kern-.1667em\lower.7ex\hbox{E}\kern-.125emX}}
    
\begin{document}
\bstctlcite{IEEEexample:BSTcontrol} % This is required to limit the maximum number of authors printed in the bibliography

\title{Evaluating System-Level Fidelity with Peaked Random Circuits}

\author{
\IEEEauthorblockN{
    Martin Brieger\IEEEauthorrefmark{1}, 
    Florian Krötz\IEEEauthorrefmark{1}, 
    Minh Chung\IEEEauthorrefmark{1,}\IEEEauthorrefmark{2},
    Dieter Kranzlmüller\IEEEauthorrefmark{1,}\IEEEauthorrefmark{2}
}

\vspace{0.3cm} % Adds breathing room between names and affiliations

\IEEEauthorblockA{\IEEEauthorrefmark{1}Ludwig-Maximilians-Universität München (LMU), Munich, Germany \\
Email: brieger@cip.ifi.lmu.de, florian.kroetz@nm.ifi.lmu.de}

\vspace{0.2cm}

\IEEEauthorblockA{\IEEEauthorrefmark{2}Leibniz Supercomputing Centre (LRZ), Garching, Germany \\
Email: \{minh.chung, dieter.kranzlmueller\}@lrz.de}
}

\maketitle

\begin{abstract}
Quantum computing is progressing beyond experimental prototypes toward cloud-based services and commercially deployed turnkey systems. Evaluating and comparing platform performance requires robust, standardized measures, yet vendors frequently report heterogeneous metrics that are not directly comparable or may not capture whole-system performance. Peaked Random Circuits (PRCs) have recently been proposed as a viable path to demonstrate quantum advantage: a quantum processor can reliably identify a single known high-probability outcome while the circuits' inherent complexity may render classical simulation infeasible. To support robust, system-level evaluation and comparison, we propose a novel benchmarking protocol based on PRCs. By successively running a set of PRCs with varying qubit counts and circuit depths, we quantify a system's ability to identify the deterministic peak despite cumulative noise, gate errors, and connectivity constraints. We demonstrate the benchmark experimentally on IQM's superconducting and AQT's trapped-ion architectures. Our results show that the PRC benchmark provides a holistic yet high-precision view of hardware performance and offers researchers and industry stakeholders a practical and scalable framework for the continuous comparison and evaluation of current and future quantum systems.
\end{abstract}


\section{Introduction} \label{sec:intro}

The rapid expansion of the quantum computing landscape has fostered a diverse ecosystem of commercial backends, spanning modalities ranging from superconducting~\cite{gambetta2017bsuperconductingqpu}, trapped-ion~\cite{bruzewicz2019trappedionqpu}, and neutral-atom~\cite{henriet2020neutralatomqpu} to photonic~\cite{browne2005opticqc} quantum computers. As hardware architectures diverge, the necessity for architecture-agnostic performance metrics becomes critical. Traditional benchmarks often struggle to capture the nuances of system-level fidelity—the holistic ability of a machine to maintain coherence and gate accuracy during complex executions—across these heterogeneous platforms.

In this paper, we propose a benchmarking methodology based on Peaked Random Circuits (PRCs). Originally conceptualized by Aaronson and Zhang to prove quantum advantage~\cite{aaronson2024verifiablequantumadvantagepeaked, zhang2025complexityprc}, PRCs are structured, quasi-random circuits designed to produce a predictable output distribution dominated by a single known ``peaked'' outcome. Provided PRCs cannot be efficiently simulated classically, a quantum machine that identifies the peaked bitstring would demonstrate a clear and easily verifiable quantum advantage.

By executing a set of PRCs with a known target bitstring, we can directly gauge system-level fidelity. We assess the device in two ways: first, by its ability to identify the correct peak, and second, by the achieved contrast between that peak and the surrounding "noise floor". To quantify these observations we introduce two evaluation metrics: Peak Identification and Fidelity Error ($\mathcal{F}$). The former maps a clear operational boundary for the hardware, while the latter yields a transparent view of how hardware noise degrades the results. 

Because they capture high-resolution noise effects, our metrics expose details that aggregate standards fail to resolve. Compared to the industry-standard performance metric, Quantum Volume (QV)~\cite{pelofske2022qvpractice}, the PRC benchmark offers three distinct advantages:

\begin{enumerate}
    \item \textit{Architectural Flexibility}: QV mandates "square" circuits (circuit depth equals the number of qubits) and consequently provides a local view of a best-quality subset of qubits. The PRC benchmark offers a heterogeneous testbed spanning the entire register size range, providing a holistic view of whole-system performance.

    \item \textit{Precision}: QV condenses system performance into a single-number metric that, while easy to compare, only reflects the system's operational ceiling. The PRC benchmark remains approachable yet captures nuanced differences in fidelity.

    \item \textit{Scalability}: QV does not scale to the circuit sizes needed to probe cutting-edge, 50+ qubit devices, as it requires exact classical simulation to identify heavy outputs. PRCs can be built without exact simulation, making the protocol robust and well-suited for more capable, next-generation hardware.
\end{enumerate}

The core contributions of this work are threefold. First, we repurpose Peaked Random Circuits (PRCs) as a system-level, architecture-agnostic benchmark, extending their original application to a universal performance-assessment framework. Second, we introduce tailored evaluation metrics that accurately capture the distinctive performance aspects of PRCs, such as peak-recovery fidelity. Third, we validate our methodology across platforms by conducting extensive experiments on two quantum computers: a superconducting system from IQM~\cite{heimonen2023iqmquantum} and a trapped-ion system from AQT~\cite{pogorelov2021compactiontrapaqt}.

The remainder of this paper is organized as follows: Section~\ref{sec:relate-work} summarizes related works and established quantum benchmarking methods. Section~\ref{sec:prc-adaptation} provides the theoretical background on PRCs and our adaptation towards a new benchmarking framework for quantum computers. Section ~\ref{sec:eval-metrics} introduces the evaluation metrics and section~\ref{sec:protocol} outlines the benchmarking protocol. In section~\ref{sec:exps-methodology}, we detail the devices' characteristics, present the experimental setup and discuss the results. Finally, in Section~\ref{sec:conclusion}, we conclude the paper and outline potential future works.

% ---------------------------------------------
% Related Work
% ---------------------------------------------
\section{Related Work} \label{sec:relate-work}

A range of benchmarking methods and frameworks has been proposed to characterize the performance of NISQ devices, each emphasizing different aspects of hardware quality~\cite{li2023qasmbench,proctor2025benchmarking,lorenz2025systematicsurvey}. 
At the system level, \emph{Quantum Volume} (QV) has become a common metric for comparing quantum processors, as it captures the largest random-square circuit a device can successfully execute under a heavy-output criterion~\cite{cross2019quantumvolume}. However, QV compresses device performance into a single scalar and may not fully reflect how structured interference patterns degrade under realistic noise and connectivity constraints~\cite{pelofske2022qvpractice}. \emph{CLOPS} complements QV by quantifying execution throughput in terms of circuit layer operations per second, but it primarily measures speed rather than computational fidelity~\cite{wack2021clops,amico2023definingbestbm}. These metrics are therefore useful for broad comparison, yet they provide only a partial view of end-to-end computational reliability. Besides, QV and CLOPS are also analyzed as the benchmarks based on the efficiency of executing a specific type of circuits. Following this, QLOPS is more clearly defined for general quantum algorithms~\cite{kong2025qlops}.

At a lower level, \emph{Randomized Benchmarking} (RB) remains a standard tool for estimating average gate errors~\cite{magesan2011rb,silva2025handsonrb}. RB is primarily used to characterize one- and two-qubit operations, but it is not designed to probe the cumulative effects of qubit mapping as well as connectivity limitations in larger multi-qubit circuits. Other approaches, such as using mirror circuits for scalable randomized benchmarking, extend this idea to deeper and wider circuits by constructing invertible circuit instances that ideally return the system to a known output state~\cite{proctor2022rbmirror}. These methods improve scalability for system-level characterization but still primarily focus on recovering the mirrored structure for flexible randomized benchmarking of Clifford gates rather than prioritizing the high-probability outcomes necessary to probe signal integrity above noise.

At the application-oriented or algorithmic level, many studies focus on benchmarking quantum hardware through reference workloads such as variational algorithms, search routines, or chemistry-inspired circuits~\cite{lubinski2023qcappbenchmark,finvzgar2022quark}. These benchmarks are valuable because they reflect end-to-end behavior in realistic use cases, but their results can depend strongly on the chosen workload, compilation strategy, and hardware-specific optimizations. 

In this paper, we use \emph{Peaked Random Circuits} (PRCs)~\cite{aaronson2024verifiablequantumadvantagepeaked, zhang2025complexityprc} to provide a more architecture-independent stress test of computational fidelity. By examining whether and "how well" a device can preserve a deterministic output peak as circuit depth and qubit count increase, PRCs directly probe the system’s ability to maintain the interference structure required for reliable quantum computation across different hardware platforms.

% ---------------------------------------------
% Simplifying PRC
% ---------------------------------------------
\section{Peaked Random Circuits for System-Level Benchmarking} \label{sec:prc-adaptation}
\subsection{Structure}

\begin{figure}[t]
\centering
    \begin{tikzpicture}[scale=0.75, transform shape]
% Parameter Definitionen
\def\numQubits{6}
\def\numLayers{10}
\def\gateWidth{0.8}
\def\gateHeight{1.5}
\def\lineSpacing{0.7}
\def\layerSpacing{0.94}

% title:
\node[anchor=base] at (5, 0.6) {\Large PRC Structure};

% Zeichne die 6 Qubit-Linien und Labels
\foreach \i in {1,...,\numQubits} {
    \draw (0, -\i*\lineSpacing) -- (\numLayers*\layerSpacing + 0.7, -\i*\lineSpacing);
    \node[left] at (0, -\i*\lineSpacing) {$\ket{0}$};
}

% layers:
\node[anchor=base] at (-0.06, -0.1) {Layer:};

% -------------------------------------------------------------
% macro that draws a rectangle and puts the supplied label in front
% -------------------------------------------------------------
\newcommand{\drawgate}[4]{%
  % #1 = layer number (\l)
  % #2 = vertical index (g)
  % #3 = colour name
  % #4 = label (empty string means no label)
  \draw[rounded corners=5pt,
        draw=#3,
        fill=#3!30]
    (#1*\layerSpacing - \gateWidth/2,
     -#2*\lineSpacing - \gateHeight/2 -0.25)
    rectangle ++(\gateWidth,\gateHeight -0.2)
    node[pos=0.5,inner sep=0]%
      {#4};
}
% -------------------------------------------------------------
\foreach \l in {1,...,\numLayers} {
    % colour selection
    \pgfmathsetmacro{\layerColor}{
        \l <= 5 ? "supergreen" : "cyan"
    }

    % pattern (odd → 3 gates, even → 2 gates, reversed after layer 5)
    \pgfmathsetmacro{\isEven}{
        (\l > 5) ?
            (mod(\l,2) == 0 ? 0 : 1)   % reverse after 5
        :
            (mod(\l,2) == 0 ? 1 : 0)   % normal
    }

    % ---- decide which label to write -------------------------
    \ifnum\l>5
        \def\mylabel{$U(\theta)$}   % second half
    \else
        \def\mylabel{$U$}        % first half
    \fi

    \ifnum\isEven=0
        % odd layer → gates 1,3,5
        \foreach \g in {1,3,5} {
            \drawgate{\l}{\g}{\layerColor}{\mylabel}
        }
    \else
        % even layer → gates 2,4
        \foreach \g in {2,4} {
            \drawgate{\l}{\g}{\layerColor}{\mylabel}
        }
    \fi

    \ifnum\l<10
        \node[anchor=base] at (\l*\layerSpacing, -0.1) {\l};
    \else
        \node[anchor=base] at (\l*\layerSpacing + 0.42, -0.1) {$10\,(=d)$};
    \fi
}

% Horizontale Klammern (Unterseite)
\draw [decorate, decoration={brace, amplitude=6pt, mirror}]
    (1*\layerSpacing - \gateWidth/2, -4.8) -- (5*\layerSpacing + \gateWidth/2, -4.8)
    node [midway, yshift=-0.6cm] {Random Half $R$};

\draw [decorate, decoration={brace, amplitude=6pt, mirror}]
    (6*\layerSpacing - \gateWidth/2, -4.8) -- (10*\layerSpacing + \gateWidth/2, -4.8)
    node [midway, yshift=-0.6cm] {Peaking Half $P(\pmb\theta)$};

\end{tikzpicture}
\caption{Mirror architecture of the Peaked Random Circuits (PRCs) we use for benchmarking. The green and blue rectangles illustrate two-qubit unitary gates. We refer to the total number of layers as the circuit's depth $d$.}
\label{fig:prc_structure}
\end{figure}
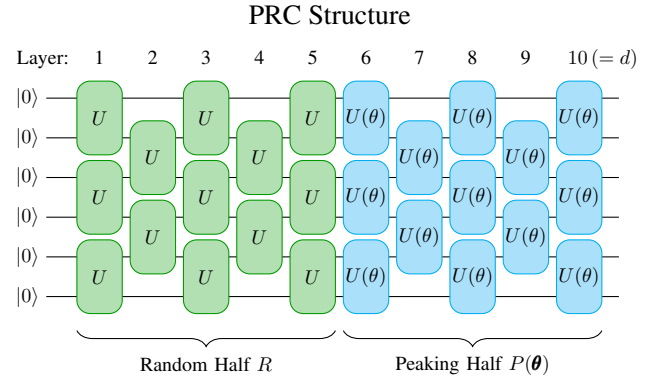

While the PRC benchmarking protocol can in principle accommodate various circuit architectures, we demonstrate it here using a mirror-circuit architecture established in the foundational PRC papers \cite{aaronson2024verifiablequantumadvantagepeaked, zhang2025complexityprc}. This approach builds on the fact that any mirrored, unitary circuit $U^{\dagger}U$ is trivially peaked at the input register. 

The construction of $n$-qubit circuits with depth $d$ consists of three steps.
First, $d/2$ layers of random two-qubit unitary gates are generated and arranged in a $1D$ brick-wall pattern, forming the "random half" of the circuit, denoted $R$. Conceptually, the random half uniformly distributes the probability mass over the state space.
Second, another $d/2$ layers of randomly generated two-qubit unitaries $P(\pmb\theta)$ are appended, resulting in a structurally mirrored, yet still random circuit $C=P(\pmb\theta)R$.
Third, the gate parameters $\pmb\theta$ are optimized to maximize the expected value of a target state  $\ket{s}$, given the all-zero input register:
\begin{equation}
    \max_{{\pmb\theta}} |\langle{s|P(\pmb\theta)R}|{0^n}\rangle|^2.
\end{equation}
The optimization forms an interference pattern that concentrates the spread probability mass, causing the target outcome $s$ to "peak", and turning $P(\pmb\theta)$ into the circuit's "peaking half". 

Figure~\ref{fig:prc_structure} illustrates the PRC architecture. The circuit's $n$ qubits are represented by horizontal lines, intersected by a sequence of two-qubit unitary gates that strictly connect their neighboring qubits, depicted as green and blue rectangles. Each layer of gates is offset from the next, forming a brick-wall pattern that ensures full entanglement across the register over time.

\begin{figure}[t]
\centering
    \input{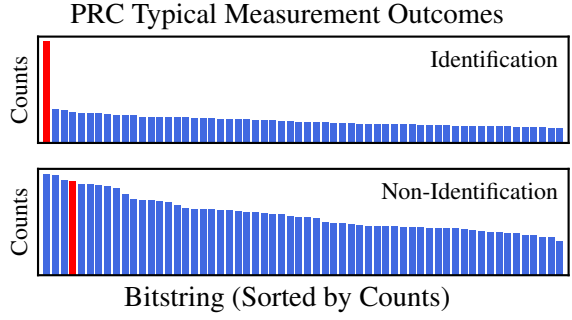}
\caption{Typical output histograms of PRCs obtained from several runs on a quantum device. The bar heights represent the frequencies of the measured bitstrings. The target bitstring is highlighted in red. In the upper panel, the most frequent result matches the target bitstring, illustrating \textit{identification}. In the lower panel, a different bitstring peaks, demonstrating \textit{non-identification}.}
\label{fig:prc_hist}
\end{figure}

When PRCs are executed on a quantum processor for a given number of shots, we expect $s$ to be the most frequent outcome. We refer to this result as \textit{identification}. Conversely, \textit{non-identification} occurs when $s$ is not the dominant sample. Figure \ref{fig:prc_hist} exemplifies the two events with typical outcome histograms.

\subsection{Generating Benchmark Circuits}
To align with the capabilities of the devices under test, we constructed a set of PRCs spanning a register size of $n\in[2,20]$ and circuit depths $d\in[2,50]$. For odd-depth circuits, we shortened the depth of the random half to $d/2-1$. We found that a more skewed random-to-peaking layer ratio either diminishes the peak probability achievable within a fixed optimization period or requires additional runtime to achieve the desired peakedness. The two-qubit circuits present an edge case: as their register size cannot accommodate the alternating brick-wall layout, their gates are aligned sequentially.

Building on the source code released with the foundational papers, we represented the circuit as a tensor network and optimized it using the quimb library in Python. A hybrid optimization scheme, comprising 5,000 iterations of the L-BFGS-B algorithm followed by 10,000 iterations of the Adam optimizer, resulted in an ideal trade-off between maximizing peak probability and maintaining computational tractability.

The default behavior of the circuits is to concentrate probability on the all-zero bitstring. Any alternative peak outcome $s$ can be realized by inserting NOT gates into the last layer and fusing them with the neighboring two-qubit unitaries. To determine whether this modification affects our performance metrics, we created a set of 10-qubit PRCs whose peaked bitstrings are multiples of $00$, $01$, $10$, and $11$. Each circuit-bitstring pair was executed five times on our target devices. The results showed no discernible variation, prompting us to keep the default.

\subsection{Difficulty} \label{subsec:difficulty-mps}

The difficulty of any quantum-computational task is fundamentally tied to the circuit size: larger registers and deeper gate sequences require higher fidelity and stronger coherence times to execute successfully. 

For PRCs, an increase in depth is accompanied by a rapid growth of entanglement originating from the brick-wall structure: every added layer of two-qubit gates entangles qubits across the entire register. Even a single-layer increment introduces new entangling bonds that will stress the devices' coherence times and gate-fidelity budgets.

\begin{figure}[t]
  \centering
    \input{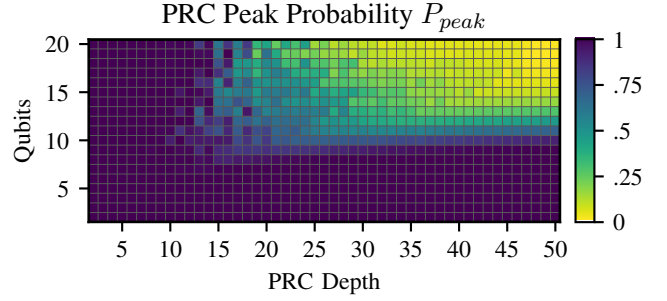}
\caption{Probability of the circuit to output the peaked, target bitstring of our set of $931$ PRCs by their number of qubits and depth. The values correspond to the results obtained in the optimization phase of circuit generation.}
\label{fig:peakedness}
\end{figure}

Figure~\ref{fig:peakedness} quantifies how this scaling manifests in the peak probability $P_{peak}$ of the benchmark circuits. As the number of qubits and the depth increase, the achieved probability mass on the peaked bitstring decreases. This relationship is intrinsic to the PRC design and results from performing the optimization in a single step. Larger circuits require optimizing more parameters, thereby increasing computational cost and reducing the attainable peak probability. 

To scale PRC generation, it has been proposed to use a discrete circuit design that sequentially optimizes independent blocks of gates \cite{udalovpaper} or embed a circuit reminiscent of the identity between the random and peaking half \cite{bluequbitpaper}. However, we favor the simple brick-wall construction of the mirror design as it yields a dense entangling pattern and a uniform gate distribution. This uniformity is beneficial for benchmarking: it guarantees that every qubit participates equally in the dynamics of execution, prevents hidden locality shortcuts, and makes the difficulty attributable to the overall size and entanglement rather than to architecture-specific factors. Moreover, even an ideal PRC would still show declining peakedness as qubit count rises, since the total probability must be distributed over an exponentially growing output space.

A major concern with decreasing $P_{peak}$ is that the benchmark becomes harder or even impossible to solve. If the difficulty of identifying the peaked bitstring were biased by varying peak probability, the benchmark would no longer reflect the intended scaling by circuit size. We therefore must ensure that circuit size is the primary driver of difficulty. There are two considerations supporting this:

\textit{1) Simulation:} The runtime required to simulate a circuit can serve as a proxy for its intrinsic difficulty, especially for PRCs that are intended to be harder to compute classically than to run on a quantum device. Leveraging a GPU-accelerated, exact/non-truncating Matrix Product State (MPS) backend, we simulated the entire benchmark suite. Identifying the peaked bitstring classically remained tractable across all instances with runtimes below one minute, including the most demanding 20-qubit, depth 50 circuit with a peak probability of only $0.41$ percent. This result suggests that small values of $P_{peak}$ by themselves do not meaningfully affect the benchmark's tractability or difficulty.

\textit{2) Peak Dominance:} A clean separation between the peak and the rest of the distribution is crucial for successful identification. We quantify Peak Dominance as:
\begin{equation}
    \mathcal{R}_{p}=\frac{P_{peak}}{P_{second}}
\end{equation}

where $P_{peak}$ and $P_{second}$ are the probabilities of the most- and second-most-likely bitstrings, with the latter being obtained from simulation data. A large $\mathcal{R}_p$ guarantees that, given enough samples, the peak will stand out to be identified even when its absolute probability is low. Tab.~\ref{tab:1} shows the $P_{peak}$ and $\mathcal{R}_p$ values of representative benchmark PRCs, sorted by circuit size. The largest PRC has a Peak Dominance of $\mathcal{R}_p=14$. In other words, the peak appears at least fourteen times more often than any competing outcome. Smaller circuits exhibit ratios in the thousands, indicating an even clearer separation. Consequently, the peaks of our circuits are pronounced enough to remain identifiable at all times, given that enough measurement samples are produced. This evidence supports the conclusion that the decreasing peak probability likely does not affect either the benchmark’s tractability or its difficulty.

As will be shown in detail, the capability ceiling of our target NISQ hardware was reached at circuit sizes with peak probabilities in the double-digit percentage range and peak dominance in the hundreds, indicating that the diminishing peak probability does not dominate the benchmark’s difficulty.

Consequently, for the PRC benchmark, the driver of difficulty is the increase in qubit count and circuit depth, i.e., the growth of entanglement, whereas the reduction in peak probability remains a secondary effect that does not threaten the task's tractability. Nevertheless, advancing towards a future benchmark standard for ever more capable hardware will demand a better scalable method for generating PRCs.

Finally, a meaningful benchmark must exhibit a deterministic, incremental increase in difficulty from one instance to the next. To achieve this, we built two reference circuits that are structurally similar to the random and peaking halves of the largest benchmark circuit. The largest PRC adapts the gates of the first reference circuit for the random half and uses the gates of the second reference circuit as a starting point for optimization. Each smaller PRC then reuses the same subsets of gates used to construct the next larger one, omitting only the additional gates introduced at higher sizes. Although each PRC has individually optimized and therefore varying gate parameters in the peaking half, the only systematic difference between successive circuits is the addition of a single qubit or layer, resulting in a marginal, predictable increase in entanglement and, by extension, computational difficulty. This design guarantees intra-circuit comparability across the entire benchmark suite.

\begin{table}[t]
\begin{center}
\caption{Selected PRC Peak Characteristics}
\label{tab:1}
\begin{tabular}{rrrrrr}
\toprule
Qubits & Depth & $P_{peak}$ & $\mathcal{R}_p$ & $N_{0.999}$ & $C_{max}$ \\
\midrule
5 & 10 & 0.9991 & 3840 & 15 & 0.9995 \\
5 & 20 & 0.9992 & 10969 & 15 & 0.9998 \\
5 & 30 & 0.9992 & 8210 & 15 & 0.9998 \\
5 & 40 & 0.9991 & 9141 & 15 & 0.9998 \\
5 & 50 & 0.9990 & 5585 & 15 & 0.9996 \\
10 & 10 & 0.9251 & 503 & 18 & 0.9960 \\
10 & 20 & 0.8398 & 557 & 22 & 0.9964 \\
10 & 30 & 0.7285 & 291 & 29 & 0.9932 \\
10 & 40 & 0.8325 & 474 & 22 & 0.9958 \\
10 & 50 & 0.9083 & 1355 & 18 & 0.9985 \\
15 & 10 & 0.9990 & 9990 & 15 & 0.9998 \\
15 & 20 & 0.5910 & 328 & 44 & 0.9939 \\
15 & 30 & 0.3382 & 338 & 133 & 0.9941 \\
15 & 40 & 0.1108 & 139 & 1238 & 0.9857 \\
15 & 50 & 0.0913 & 152 & 1823 & 0.9869 \\
20 & 10 & 0.9877 & 898 & 16 & 0.9978 \\
20 & 20 & 0.4170 & 199 & 87 & 0.9900 \\
20 & 30 & 0.0936 & 234 & 1737 & 0.9915 \\
20 & 40 & 0.0499 & 166 & 6095 & 0.9881 \\
20 & 50 & 0.0041 & 14 & 890634 & 0.8646 \\
\bottomrule
\end{tabular}
\end{center}
\end{table}

\subsection{Sampling Strategy}

To statistically distinguish the probability of the target outcome $P_{peak}$ from the background distribution, we assume a uniform depolarizing noise model over the remaining Hilbert space. For an $n$-qubit system with $2^n$ computational basis states, the background probability $P_{bg}$ for any non-target state is roughly uniformly distributed as $P_{bg} = (1 - P_{peak}) / (2^n - 1)$. 

Using the Hoeffding bound, the number of measurement shots $N$ required to identify the target state as the distribution's peak with a confidence level of $1 - \delta$ is given by:
\begin{equation}
    N_{1-\delta} \geq \frac{2}{\left( P_{peak} - \frac{1 - P_{peak}}{2^n - 1} \right)^2} \ln\left(\frac{2}{\delta}\right).
\end{equation}

For a sufficiently large number of qubits $n$, the background probability $P_{bg}$ becomes exponentially suppressed. Thus, the required sampling complexity scales asymptotically as $N = \mathcal{O}(1/P_{peak}^2)$, reflecting the inherent overhead of resolving the spectral gap in the probability distribution.

The fifth column of Tab.~\ref{tab:1} presents the theoretical minimum number of shots, $N_{0.999}$, required to identify the peak with 99.9 percent confidence. For most benchmark circuits, the required sampling budget remains low in the double digits.

%-------------------------------------------------------------------------------

\section{Evaluation Metrics} \label{sec:eval-metrics}

The PRC benchmark is designed to progress from high-level system stress testing to granular fidelity analysis. Unlike aggregate statistical measures, our metrics focus on the "sharpness" of the result: the machine's ability to distinguish a specific computational signal from the background noise floor. We employ a dual-metric approach that captures both the ordinal correctness (Was the bitstring recovered?) and the computational margin (How well was it recovered?).

\subsection{Success Criteria: Peak Identification}
To start with and to obtain a clear operational boundary for the hardware, we apply a binary Peak Identification check. In the event of identification, i.e., the target bitstring $s$ being the most frequent outcome state in the experimental distribution, the test is passed. In the event of non-identification, i.e., the dominant outcome not matching $s$, the test fails.

\subsection{Quantifying Sharpness: Relative Peakedness ($C$)}

To measure "how well" a device is able to recover the peaked bitstring in case of a successful identification, we propose a Relative Peakedness metric $C_{exp}\in[0,1]$, calculated as the normalized difference:
\begin{equation}
    C_{exp}=\frac{\hat{P}_{peak}-\hat{P}_{second}}{\hat{P}_{peak}+\hat{P}_{second}}.
\end{equation}
$\hat{P}_{peak}$ corresponds to the peak probability and $\hat{P}_{second}$ to the probability of the second most probable outcome, both obtained from the experimental distribution. $C = 1$ indicates a perfectly "sharp" peak that is clearly distinguishable from the strongest competing error state, whereas $C = 0$ indicates that the hardware can no longer differentiate the signal from the background noise.

Our metric corrects for the fact that the state space expands as $2^n$ with increasing qubit count $n$, causing the probability of any one basis state to vanish, including the peak. As $\hat{P}_{peak}$ naturally falls with a growing register size, larger circuits could be penalized, regardless of hardware quality. The normalization ensures that the benchmark is a fair test of constructive interference rather than a measure of state-space sparsity.

\subsection{Fidelity Error ($\mathcal{F}$)}
In practice, the benchmark circuits exhibit non-uniform peak characteristics that must be accounted for. Given an ideal device, a perfectly peaked PRC with $P_{peak}=1$ would yield measurements consistent with $C_{exp}=1$. By contrast, an imperfect PRC with $P_{peak}\ll1$ and low Peak Dominance $\mathcal{R}_p$ would produce outcomes that, at best, reflect the circuit’s weak peak profile rather than the device’s true capability. Consequently, a peak-recovery metric must both accurately capture a device's ability to preserve signal integrity above noise and remain comparable across heterogeneous circuits. 

We achieve this by normalizing $C_{exp}$ with the best-case attainable value:
\begin{equation}
    C_{max}=\frac{P_{peak}-P_{second}}{P_{peak}+P_{second}}.
\end{equation}
$P_{peak}$ and $P_{second}$ correspond to the circuit's inherent first- and second-peak probabilities, used also to construct $\mathcal{R}_p$. The latter can be obtained either via classical simulation or directly from circuit generation.

Consequently, we define our main metric, the Fidelity Error $\mathcal{F}\in[0,1]$ by scaling $C_{exp}$ with $C_{max}$ and subtracting from one:
\begin{equation}
    \mathcal{F} = 1-\frac{C_{exp}}{C_{max}}.
\end{equation}

The ratio of the experimentally achieved to the best-case achievable Relative Peakedness quantifies how precisely a device reproduces a PRC’s intrinsic peak profile. By taking the complement of the ratio, the metric becomes a measure of "fidelity error": $\mathcal{F}=0$ indicates a perfect recovery of the PRC's peak characteristics, while $\mathcal{F}=1$ indicates a failure to recover the peak characteristic.

To provide intuition about typical peak profiles and the scale of correction required in practice, Tab.~\ref{tab:1} reports the $C_{max}$ values of representative benchmark PRCs. Although larger circuits typically show lower $P_{peak}$ and $\mathcal{R}_p$, the $C_{max}$ values used for correction remain close to one for most instances, deviating only for the largest circuit. This suggests that even relatively large PRCs retain enough peak sharpness for the device to recover, making the correction primarily a technical refinement rather than an essential adjustment. Under this interpretation, $\mathcal{F}$ reflects the hardware’s ability to differentiate the correct outcome from the most likely error state, thus providing a measure of system-level fidelity.

\section{PRC Benchmarking Protocol} \label{sec:protocol}
\begin{figure}[t]
\centering
\begin{tikzpicture}[node distance = 0.5cm and 2.5cm]

\node [data, fill=supergreen!30, align=center] (start) {Prepare a set \\ of $n \times d$ PRCs};
\node [terminator, fill=cyan!30, below=of start] (transpile) {Transpile PRC for target hardware};
\node [terminator, fill=cyan!30,  align=center, below=of transpile] (execute) {Execute PRC and \\ collect bitstrings};
\node [decision, fill=yellow!30, below=of execute, align=center] (decision) {Peak \\ identified?};
\node [terminator, fill=cyan!30, below=of decision] (compute) {Compute Fidelity Error $\mathcal{F}$};
\node [data, fill=supergreen!30, below=of compute, align=center] (plot) {Plot $n \times d$ matrix \\ and add $\mathcal{F}$-values};

\path [connector] (start) -- (transpile);
\path [connector] (transpile) -- (execute);
\path [connector] (execute) -- (decision);
\path [connector] (decision) -- (compute);
\path [connector] (compute) -- (plot);

\path [connector] (compute.east) -- ++(1.5,0) |- (transpile.east) node[right, pos=0.25, align=left, xshift=0.1cm] {Loop over all \\ $n \times d$ instances};

\coordinate (intersection) at ([xshift=1.5cm]compute.east |- decision);
\path [connector] (decision.east) -- (intersection);

\path [connector] (decision) -- node[left, pos=0.4] {Yes} (compute);
\path [connector] (decision.east) -- node[above] {No} (intersection);

\end{tikzpicture}
\caption{Flowchart of the PRC benchmarking protocol.}
\label{fig:prc_protocol}
\end{figure}

Fig.~\ref{fig:prc_protocol} illustrates the PRC benchmarking protocol. First, a set of $n \times d$ PRCs is prepared that aligns with the capabilities of the system under test: $n$ must equal the number of active qubits and $d$ should be large enough that the Peak Identification check fails at some depth within the span for every value of $n$. Comparability of the results requires that the same set of circuits is (re-) used or circuits are generated deterministically, using the same PRC architecture. The protocol proceeds iteratively over all ($n,d$) instances, starting with the smallest $n$ and exploring all depths for that register size before advancing to a larger $n$. 

Each circuit is transpiled to the system’s native gate set, with the transpiler viewed as part of the system and classical optimizations enabled. Subsequently, the circuit is executed with the number of measurement samples matching or exceeding the minimum required to identify the target state at a given confidence level.

After each run, the experimental distribution is analyzed. If the Peak Identification check fails the loop proceeds to the next instance. If it succeeds, $\mathcal{F}$ is computed and logged before handling the next instance. To conserve compute resources, remaining instances at a given qubit count can be skipped when the preceding depths (e.g., the last four) produced non-identifications. Finally, the collected data points are plotted in a $n \times d$ result lattice.

\section{Experiments and Results} \label{sec:exps-methodology}
To demonstrate the PRC benchmarking protocol experimentally, we performed a cross-platform study using two quantum systems with distinct physical modalities hosted at the Leibniz Supercomputing Centre (LRZ)~\footnote{\url{https://www.lrz.de/en/technologies/quantum-computing}} of the Bavarian Academy of Sciences. 

\subsection{Devices}

The first system is a "Radiance" superconducting device from IQM, codenamed \textbf{QExa20}, comprising $20$ qubits arranged in a square-lattice connectivity topology~\cite{heimonen2023iqmquantum}. IQM advertises a typical QV of $\ge32$ and provides the median fidelities and durations for single- and two-qubit gates, the median readout fidelity, and the maximum number of qubits that can be placed in a GHZ state with fidelity $>0.5$.

The second system is a "Marmot" trapped-ion device from AQT, codenamed \textbf{AQT20}, also with 20 qubits but featuring all-to-all connectivity~\cite{pogorelov2021compactiontrapaqt}. During our experiments, 16 of its qubits were functional. The device is promoted with a QV of $128$, and AQT reports average single- and two-qubit error rates on an 11-qubit register together with nearest-neighbor crosstalk.

Relying on the vendor-provided specifications makes it difficult to evaluate and compare the true performance of either system: aside from QV, the given metrics are not directly comparable. While the higher QV score may indicate an advantage for the AQT system, we will show that the PRC benchmark yields a more nuanced assessment.

\subsection{Execution Pipeline}

We first converted the set of circuits from their original tensor network representation to the OpenQASM format. To guarantee a deterministic, fair benchmark across the two devices, we employed the following execution protocol:

\textit{1) Transpilation:} Each ($n,d$) instance was transpiled to the native gate set of the target device using the Qiskit transpiler~\cite{javadi2024qiskitquantum}, applying the appropriate translation and scheduling passes. We enabled the highest degree of classical optimization, though, as expected, this did not reduce the number of resulting native gates and therefore did not degrade the difficulty of the benchmark tasks.

\textit{2) Software Stack:} We interfaced with the LRZ quantum devices, submitted jobs, and monitored their execution using the Qiskit adapter provided by the Munich Quantum Software Stack (MQSS)~\cite{mqss}.

\textit{3) Sampling:} We set the number of measurement samples to be produced by both devices to numbers above the circuit-specific minima required for the peak to be identified at the 0.999 confidence level.

\textit{4) Execution:} We implemented an adaptive execution strategy, skipping the remaining instances at a given register size if the last five consecutive runs produced non-identifications. Typical job runtimes ranged between 5 and 300 seconds on both devices, depending on queue latency and device calibration cycles.

\textit{5) Averaging:} To account for drift accumulated since the device's last calibration and general run-to-run variability, we performed the protocol a total of five times and averaged the resulting metrics. Although optional, this practice is advisable when devices are heavily loaded or shared, or when the interval since the last calibration differs between jobs.

\subsection{Visualizing the Results}

\begin{figure}[t]
\centering
\input{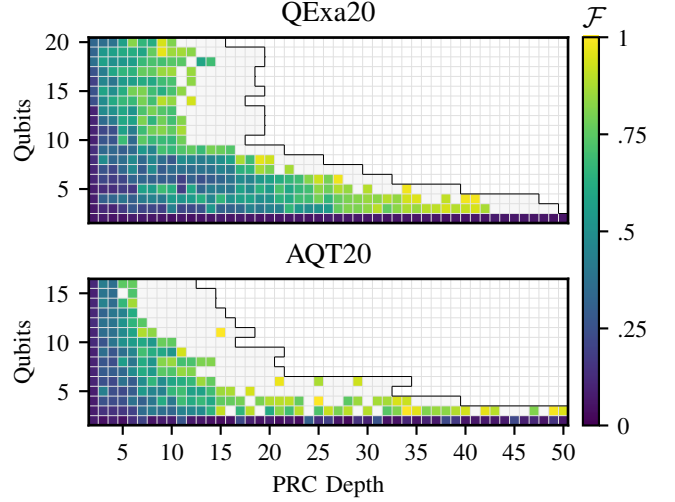}
\caption{Successful recoveries of the peaked bitstring by the circuits' depth and number of qubits for the QExa20 superconducting device and the AQT20 ion-trap device. Colored cells indicate a matching peaked bitstring in at least three of five runs. Light-gray cells indicate non-identifications. White cells indicate skipped runs resulting from previously run, lower-depth circuits yielding non-identifications. The black line separates executed circuits from skipped ones. The colormap indicates the Fidelity Error $\mathcal{F}$.}
\label{fig:fidelity_gradient}
\end{figure}

Fig.~\ref{fig:fidelity_gradient} shows the benchmark outcomes for both the Peak Identification and Fidelity Error metrics. QExa20 results are displayed in the top panel and AQT20 results in the bottom panel. Each plot is arranged as a two-dimensional lattice whose vertical axis correspond to the number of qubits $n$ and horizontal axis to circuit depth $d$. Every cell on the $n\times d$ lattice is color-coded: a colored cell signals a successful peak identification, a light-gray cell indicates a failure to identify the peak, and a white cell marks a configuration that was not executed. A black line traverses the plots, separating the circuits that were actually run (left) from those that were omitted (right). For the successful runs, the colormap reflects the Fidelity Error $\mathcal{F}$: a value close to zero means the target bitstring stands out clearly against the noise, whereas a value approaching one signals that the hardware can barely distinguish the peak from the dominant error state.

Jointly presenting the result matrices of multiple devices side by side transforms the PRC benchmark into a versatile, heterogeneous testbed. This format makes it easy to contrast systems built on different physical modalities, while the x- and y-axes can be freely adjusted to reflect each platform’s performance range and to streamline the comparison.

\subsection{Operational Boundary}

Comparing the colored cells in Fig.~\ref{fig:fidelity_gradient} to the white and gray background displays a clear operational boundary. Both devices could reliably detect the peak for modest-size circuits, yet QExa20 demonstrated a noticeably larger operational range: across the 5- to 16-qubit span, it consistently outperformed AQT20 in handling deeper PRCs. 

The shape of the boundary reflects performance scaling. As circuit width increases, the maximum attainable depth generally declines, producing a diagonal contour. Nevertheless, QExa20 scales more favorably for larger registers, continuously supporting depth-10 circuits throughout the 10- to 20-qubit region.

The distribution of successful identifications serves as a proxy for performance consistency. Across the lattice, AQT20’s successes appear more erratic—occasionally an identification success follows a series of failures, and vice versa, indicating less stable performance. In contrast, QExa20’s identifications develop mostly uniformly until the operational boundary is reached.

When binary success/failure outcomes are considered without regard to peak identification quality, QExa20 demonstrates higher overall performance, superior scaling, and greater consistency than AQT20.

\subsection{Fidelity Error Analysis}

The Fidelity Error $\mathcal{F}$, visualized by the color scale in Fig.~\ref{fig:fidelity_gradient}, enables a deep diagnostic of the success matrices, providing a more granular view of hardware performance. By examining the transition zones where $\mathcal{F}$ begins to climb from $0$ to $1$, we can distinguish between performance loss caused by adding more qubits (width) versus adding more gates (depth). This transition from a general "capability horizon" to a specific sensitivity analysis reveals that the breakdown of signal integrity is not uniform across different devices.

In both systems, the hardware's ability to differentiate the target signal from background noise vanished as the computational volume ($n \times d$) increased. However, the error rates were platform-dependent and heterogeneous, motivating a more detailed comparison. 

\begin{figure}[t]
\centering
\input{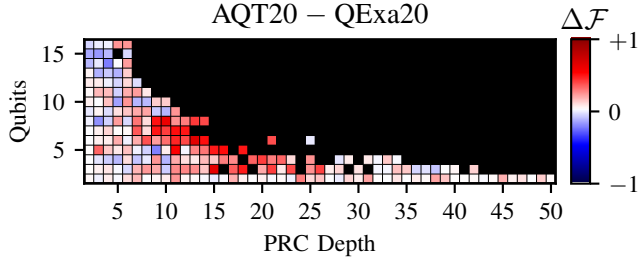}
\caption{Difference in the quality of peak identification between the AQT20 ion-trap device and the QExa20 superconducting device by the circuits' depth and number of qubits. Colored cells indicate successful peak identifications of both devices. The colormap quantifies the difference in the Fidelity Error between the two devices: $\Delta\mathcal{F}=\mathcal{F}_{AQT20}-\mathcal{F}_{QExa20}$}
\label{fig:fidelity_diff}
\end{figure}

Fig.~\ref{fig:fidelity_diff} visualizes the difference in achieved Fidelity Error between the two devices, calculated as:
\begin{equation}
    \Delta\mathcal{F}=\mathcal{F}_{AQT20}-\mathcal{F}_{QExa20}.
\end{equation}
The data values were linearly mapped to a diverging colormap defined on the symmetric interval \([-1, 1]\), centered at zero. Blue-colored cells denote configurations where AQT20 outperforms QExa20 (lower error), while red-colored cells mark configurations where QExa20 is superior (higher error for AQT20). White cells indicate equal performance. The axes remain consistent with the previous analysis. 

The presence of local red and blue “heat zones” on the lattice clearly shows that the devices' performance scales differently with circuit width and depth. QExa20 outperforms the competitor in deep circuits with fewer qubits, achieving a higher peak separation for those configurations. In contrast, AQT20 excels with broader, shallower circuits, exhibiting a lower Fidelity Error than QExa20. However, this advantage is less striking, as reflected in fewer blue cells and lighter color saturation.

This analysis highlights the strength of our dual-metric approach. Although the operational boundary alone implies that QExa20 is categorically superior than AQT20, the Fidelity Error metric reveals a more nuanced picture: AQT20 offers a small yet measurable benefit in signal preservation for some larger-register circuits with fewer gates.

\subsection{Cross-Platform Comparison}

Several observations emerge when evaluating performance across two distinct architectures:

\textit{1) Connectivity and Topology:} All-to-all qubit connectivity can be beneficial for circuits that need long-range interactions, since it removes the need for additional routing gates that would deplete the error budget more rapidly. Our benchmark PRCs are built in a brick-wall pattern with each gate strictly coupling adjacent qubits; therefore, a chain connectivity is sufficient to execute the circuits without extra routing. The QExa20’s square-lattice topology can form such a chain, so it does not incur the penalty over the fully connected qubits of AQT20. We confirmed no connectivity bias exists by performing a post-transpilation gate-count analysis: both devices executed exactly the same number of two-qubit gates for each circuit. In general, our PRC benchmark remains platform-agnostic as long as the hardware can realize a linear chain topology.

\textit{2) Hardware-Specific Noise Pathologies:} On the QExa20 system, we observed signs of crosstalk in larger registers. As more qubits were utilized, the noise floor became increasingly non-uniform, with "ghost peaks" appearing near the target bitstring, likely due to localized coherent errors. AQT20 did not exhibit this effect, offering a plausible reason for its superior performance on some of those circuits.

\textit{3) Physical Limits:} AQT20 encountered a physical limit at the $n15d15$ configuration, corresponding to 2000 native gates, beyond which the system returned gate-count errors. This limitation originates from the need to preserve stability in long ion chains and the duration of gate operations: AQT20's performance horizon is hard-capped by the total number of operations the system can maintain before the chain requires re-cooling. In practice, the operational boundary was reached before the physical limit, so the performance assessment was not impaired.

\textit{4) Modes of Failure:} Our experiments highlight two distinct failure modes in NISQ hardware. While our MPS simulations (discussed in Subsection~\ref{subsec:difficulty-mps}) recovered the peak for the largest $20$-qubit, depth-$50$ circuit in under a minute, the devices faced a "precision vs. reach" trade-off. For QExa20, the challenge is preventing the interference patterns from being buried by cumulative decoherence (precision). For AQT20, the challenge is maintaining the count of successful outcomes (reach).

\section{Conclusion} \label{sec:conclusion}

We introduced a novel benchmarking framework based on Peaked Random Circuits (PRCs). It evaluates system performance using a dual-metric approach: first, by verifying whether the target bitstring was correctly identified, and second, by quantifying the quality of the identification through the Fidelity Error ($\mathcal{F}$). This strategy yields a high-resolution view of how hardware noise destroys constructive interference, offering deeper insights than standard aggregate metrics like Quantum Volume.

Our experimental results on IQM's superconducting and AQT's trapped-ion systems reveal a clear capability horizon for current NISQ devices. While both platforms successfully identified peaks for small- to medium-scale circuits, signal integrity rapidly decayed as entanglement increased with depth and qubit count. 

Notably, our findings uncover a significant gap between the requirements for a verifiable quantum advantage demonstration and the operational limits of current hardware. For the PRC architecture used in this paper, the computational demands of maintaining a sharp peak across a $20$-qubit brick-wall structure proved to be a formidable stress test, often exceeding the coherence times and gate-fidelity budgets of state-of-the-art machines. Consequently, this work provides a practical reality check: PRCs designed for quantum advantage are currently too complex for NISQ hardware at scale. The issue is not the circuit design itself, but rather how hardware noise "washes out" the interference patterns these algorithms require.

Future work has to focus on scaling PRC generation. While our current optimization step works for small qubit counts, larger circuits will require more efficient methods, such as block-based optimization. We also plan to explore "super-peaked" circuits, which trade randomness for a stronger signal to better test the limits of next-generation processors.

\section*{Acknowledgments}

The authors gratefully acknowledge the use of the quantum systems QExa20 and AQT20, operated by the Leibniz Supercomputing Centre (LRZ) in Garching, Germany, providing the computational resources for this work.

Parts of this work were funded by the Munich Quantum Valley (MQV) initiative and the Q-DESSI project. Hardware operations and associated software stack projects were supported by the German Federal Ministry of Research, Technology and Space (BMFTR) and the Bavarian State Ministry of Science and the Arts (StMWK).

\bibliographystyle{IEEEtran}
\bibliography{prc.bib}

@IEEEtranBSTCTL{IEEEexample:BSTcontrol,
  CTLuse_forced_etal       = "yes",
  CTLmax_names_forced_etal = "6",
  CTLnames_show_etal       = "5" 
}

@article{browne2005opticqc,
  title={Resource-efficient linear optical quantum computation},
  author={Browne, Daniel E and Rudolph, Terry},
  journal={Physical Review Letters},
  volume={95},
  number={1},
  pages={010501},
  year={2005},
  publisher={APS},
  url={https://doi.org/10.1103/PhysRevLett.95.010501}
}

@article{magesan2011rb,
   title={Scalable and Robust Randomized Benchmarking of Quantum Processes},
   volume={106},
   ISSN={1079-7114},
   url={http://dx.doi.org/10.1103/PhysRevLett.106.180504},
   doi={10.1103/physrevlett.106.180504},
   number={18},
   journal={Physical Review Letters},
   publisher={American Physical Society (APS)},
   author={Magesan, Easwar and Gambetta, J. M. and Emerson, Joseph},
   year={2011},
   month=may
}

@article{gambetta2017bsuperconductingqpu,
  title={Building logical qubits in a superconducting quantum computing system},
  author={Gambetta, Jay M and Chow, Jerry M and Steffen, Matthias},
  journal={npj quantum information},
  volume={3},
  number={1},
  pages={2},
  year={2017},
  publisher={Nature Publishing Group UK London},
  url={https://doi.org/10.1038/s41534-016-0004-0}
}

@article{bruzewicz2019trappedionqpu,
  title={Trapped-ion quantum computing: Progress and challenges},
  author={Bruzewicz, Colin D and Chiaverini, John and McConnell, Robert and Sage, Jeremy M},
  journal={Applied physics reviews},
  volume={6},
  number={2},
  year={2019},
  publisher={AIP Publishing},
  url={https://doi.org/10.1063/1.5088164}
}

@article{cross2019quantumvolume,
  title={Validating quantum computers using randomized model circuits},
  author={Cross, Andrew W and Bishop, Lev S and Sheldon, Sarah and Nation, Paul D and Gambetta, Jay M},
  journal={Physical Review A},
  volume={100},
  number={3},
  pages={032328},
  year={2019},
  publisher={APS},
  url={https://doi.org/10.1103/PhysRevA.100.032328}
}

@article{henriet2020neutralatomqpu,
  title={Quantum computing with neutral atoms},
  author={Henriet, Lo{\"\i}c and Beguin, Lucas and Signoles, Adrien and Lahaye, Thierry and Browaeys, Antoine and Reymond, Georges-Olivier and Jurczak, Christophe},
  journal={Quantum},
  volume={4},
  pages={327},
  year={2020},
  publisher={Verein zur F{\"o}rderung des Open Access Publizierens in den Quantenwissenschaften},
  url={https://doi.org/10.22331/q-2020-09-21-327}
}

@article{wack2021clops,
  title={Quality, Speed, and Scale: Three key attributes to measure the performance of near-term quantum computers},
  author={Wack, Andrew and Paik, Hanhee and Javadi-Abhari, Ali and Jurcevic, Petar and Faro, Ismael and Gambetta, Jay M and Johnson, Blake R},
  journal={arXiv preprint arXiv:2110.14108},
  volume={2110},
  year={2021},
  url={https://doi.org/10.48550/arXiv.2110.14108}
}

@article{pogorelov2021compactiontrapaqt,
  title={Compact ion-trap quantum computing demonstrator},
  author={Pogorelov, Ivan and Feldker, Thomas and Marciniak, Ch D and Postler, Lukas and Jacob, Georg and Krieglsteiner, Oliver and Podlesnic, Verena and Meth, Michael and Negnevitsky, Vlad and Stadler, Martin and others},
  journal={PRX quantum},
  volume={2},
  number={2},
  pages={020343},
  year={2021},
  publisher={APS},
  url={https://doi.org/10.1103/PRXQuantum.2.020343}
}

@article{proctor2022rbmirror,
   title={Scalable Randomized Benchmarking of Quantum Computers Using Mirror Circuits},
   volume={129},
   ISSN={1079-7114},
   url={http://dx.doi.org/10.1103/PhysRevLett.129.150502},
   DOI={10.1103/physrevlett.129.150502},
   number={15},
   journal={Physical Review Letters},
   publisher={American Physical Society (APS)},
   author={Proctor, Timothy and Seritan, Stefan and Rudinger, Kenneth and Nielsen, Erik and Blume-Kohout, Robin and Young, Kevin},
   year={2022},
   month=oct
}

@article{pelofske2022qvpractice,
  title={Quantum volume in practice: What users can expect from {NISQ} devices},
  author={Pelofske, Elijah and B{\"a}rtschi, Andreas and Eidenbenz, Stephan},
  journal={IEEE Transactions on Quantum Engineering},
  volume={3},
  pages={1--19},
  year={2022},
  publisher={IEEE},
  url={https://doi.org/10.1109/TQE.2022.3184764}
}

@inproceedings{finvzgar2022quark,
  author={Finžgar, Jernej Rudi and Ross, Philipp and Hölscher, Leonhard and Klepsch, Johannes and Luckow, Andre},
  booktitle={2022 IEEE International Conference on Quantum Computing and Engineering (QCE)}, 
  title={QUARK: A Framework for Quantum Computing Application Benchmarking}, 
  year={2022},
  volume={},
  number={},
  pages={226-237},
  doi={10.1109/QCE53715.2022.00042}
}

@inproceedings{amico2023definingbestbm,
  title={Defining best practices for quantum benchmarks},
  author={Amico, Mirko and Zhang, Helena and Jurcevic, Petar and Bishop, Lev S and Nation, Paul and Wack, Andrew and McKay, David C},
  booktitle={2023 IEEE International Conference on Quantum Computing and Engineering (QCE)},
  volume={1},
  pages={692--702},
  year={2023},
  organization={IEEE},
  url={https://doi.org/10.1109/QCE57702.2023.00084}
}

@article{li2023qasmbench,
  title={Qasmbench: A low-level quantum benchmark suite for {NISQ} evaluation and simulation},
  author={Li, Ang and Stein, Samuel and Krishnamoorthy, Sriram and Ang, James},
  journal={ACM Transactions on Quantum Computing},
  volume={4},
  number={2},
  pages={1--26},
  year={2023},
  publisher={ACM New York, NY},
  url={https://doi.org/10.1145/3550488}
}

@article{lubinski2023qcappbenchmark,
  title={Application-oriented performance benchmarks for quantum computing},
  author={Lubinski, Thomas and Johri, Sonika and Varosy, Paul and Coleman, Jeremiah and Zhao, Luning and Necaise, Jason and Baldwin, Charles H and Mayer, Karl and Proctor, Timothy},
  journal={IEEE Transactions on Quantum Engineering},
  volume={4},
  pages={1--32},
  year={2023},
  publisher={IEEE},
  url={https://doi.org/10.1109/TQE.2023.3253761}
}

@incollection{heimonen2023iqmquantum,
  title={Quantum Computing at {IQM}},
  author={Heimonen, Hermanni and Auer, Adrian and Bergholm, Ville and de Vega, In{\'e}s and M{\"o}tt{\"o}nen, Mikko},
  booktitle={Impact of Scientific Computing on Science and Society},
  pages={373--393},
  year={2023},
  publisher={Springer},
  url={https://doi.org/10.1007/978-3-031-29082-4_22}
}

@article{aaronson2024verifiablequantumadvantagepeaked,
  title={On verifiable quantum advantage with peaked circuit sampling}, 
  author={Scott Aaronson and Yuxuan Zhang},
  year={2024},
  journal={arXiv preprint arXiv:2404.14493},
  url={https://doi.org/10.48550/arXiv.2404.14493}, 
}

@article{javadi2024qiskitquantum,
  title={Quantum computing with Qiskit},
  author={Javadi-Abhari, Ali and Treinish, Matthew and Krsulich, Kevin and Wood, Christopher J and Lishman, Jake and Gacon, Julien and Martiel, Simon and Nation, Paul D and Bishop, Lev S and Cross, Andrew W and others},
  journal={arXiv preprint arXiv:2405.08810},
  url={https://doi.org/10.48550/arXiv.2405.08810}, 
  year={2024}
}

@article{zhang2025complexityprc,
  title={Complexity and hardness of random peaked circuits},
  author={Zhang, Yuxuan},
  journal={arXiv preprint arXiv:2510.00132},
  year={2025},
  url={https://doi.org/10.48550/arXiv.2510.00132}
}

@article{kong2025qlops,
  title={Benchmarking fault-tolerant quantum computing hardware via {QLOPS}},
  author={Kong, Linghang and Zhang, Fang and Chen, Jianxin},
  journal={ACM Transactions on Quantum Computing},
  year={2025},
  publisher={ACM New York, NY},
  url={https://doi.org/10.1145/3797968}
}

@article{lorenz2025systematicsurvey,
  title={Systematic benchmarking of quantum computers: Status and recommendations},
  author={Lorenz, Jeanette Miriam and Monz, Thomas and Eisert, Jens and Reitzner, Daniel and Schopfer, F{\'e}licien and Barbaresco, Fr{\'e}d{\'e}ric and Kurowski, Krzysztof and van der Schoot, Ward and Strohm, Thomas and Senellart, Jean and others},
  journal={arXiv preprint arXiv:2503.04905},
  year={2025},
  url={https://doi.org/10.48550/arXiv.2503.04905}
}

@article{proctor2025benchmarking,
  title={Benchmarking quantum computers},
  author={Proctor, Timothy and Young, Kevin and Baczewski, Andrew D and Blume-Kohout, Robin},
  journal={Nature Reviews Physics},
  volume={7},
  number={2},
  pages={105--118},
  year={2025},
  publisher={Nature Publishing Group UK London},
  url={https://doi.org/10.1038/s42254-024-00796-z}
}

@article{silva2025handsonrb,
   title={Hands-on introduction to randomized benchmarking},
   ISSN={2590-1990},
   url={http://dx.doi.org/10.21468/SciPostPhysLectNotes.97},
   DOI={10.21468/scipostphyslectnotes.97},
   journal={SciPost Physics Lecture Notes},
   publisher={Stichting SciPost},
   author={Silva, Ana and Greplova, Eliska},
   year={2025},
   month=jul
}

@article{mqss,
  title        = {The Munich Quantum Software Stack: Connecting End Users, Integrating Diverse Quantum Technologies, Accelerating {HPC}},
  shorttitle   = {{The Munich Quantum Software Stack}},
  author       = {Burgholzer, Lukas and Echavarria, Jorge and Hopf, Patrick and Stade, Yannick and Rovara, Damian and Schmid, Ludwig and Kaya, Ercüment and Mete, Burak and Farooqi, Muhammad Nufail and Chung, Minh and De Pascale, Marco and Schulz, Laura and Schulz, Martin and Wille, Robert},
  year         = 2025,
  journal={arXiv preprint arXiv:2509.02674},
  url={https://doi.org/10.48550/arXiv.2509.02674},
}

@misc{bluequbitpaper,
      title={Heuristic Quantum Advantage with Peaked Circuits}, 
      author={Hrant Gharibyan and Mohammed Zuhair Mullath and Nicholas E. Sherman and Vincent P. Su and Hayk Tepanyan and Yuxuan Zhang},
      year={2025},
      journal={arXiv preprint arXiv:2510.25838},
      url={https://doi.org/10.48550/arXiv.2510.25838}, 
}

@article{udalovpaper,
      title={A Method for Constructing Quasi-Random Peaked Quantum Circuits}, 
      author={O. G. Udalov},
      year={2025},
      journal={arXiv preprint arXiv:2508.07491},
      url={https://doi.org/10.48550/arXiv.2508.07491}, 
}

\end{document}